\documentstyle[twocolumn,aps,epsfig]{revtex}

\newcommand{\gsim}{\mathrel{\raisebox{-.6ex}{$\stackrel{\textstyle>}{\sim}$}}}

\begin{document}
\draft

\title{
On Black Hole Detection with the OWL/Airwatch Telescope 
}
\author{Sharada Iyer Dutta$^1$, Mary Hall Reno$^{2}$ and Ina Sarcevic$^{3,4}$}
\address{
$^1$Department of Physics, SUNY Stony Brook, Stony Brook, NY 
11794\\
$^2$Department of Physics and Astronomy, University of Iowa, Iowa City,
Iowa 52242\\
$^3$Department of Physics, University of Arizona, Tucson, Arizona
85721\\
$^4$Theory Division, CERN, CH-1211 Geneva 23, Switzerland\\
}

\wideabs{
\maketitle

\begin{abstract}
\widetext 
In scenarios with large extra dimensions and TeV scale gravity 
ultrahigh energy neutrinos produce black holes in their interactions 
with the nucleons.  We show that ICECUBE and OWL may 
observe large number of black hole events and provide valuable 
information about the fundamental Planck scale and the number of 
extra dimensions. OWL is especially well suited to observe black hole
events produced by neutrinos from the interactions of cosmic rays
with the 3 K background radiation. Depending on the parameters of
the scenario of large extra dimensions and on the
flux model, as many as 28 events per year
are expected for a Planck scale of 3 TeV. 

\end{abstract}
}
\vskip 0.1true in

\narrowtext
\section{Introduction}

A recent proposal of lowering the fundamental Planck scale to the TeV range
has provided a new perspective on studying black hole formation in
ultra-relativistic collisions \cite{Antoniadis:1990ew}.  It has been
argued that in particle collisions with energies above the Planck
scale $M_D$ ($M_D \sim$ TeV), black holes can be produced and their
production and decay can be described semiclassically and
thermodynamically \cite{Emparan:2000rs}.  In proton-proton collisions
at the Large Hadron Collider (LHC) at CERN with center of mass energy
of several TeV, for example, the distinctive characteristics of black
hole production would be large multiplicity events
\cite{Giddings:2001bu,Rizzo:2001dk,Ringwald:2001vk}.  
The event rates depend strongly on the ratio
of the minimum mass of the black hole and the Planck scale and to a
lesser extent on the number of extra dimensions \cite{Rizzo:2001dk}.
Recently it has also been pointed out that cosmic ray detectors
sensitive to neutrino induced air showers, such
as large Pierre Auger Observatory, could detect black holes
produced in the neutrino interactions with the atmosphere 
\cite{Ringwald:2001vk,Emparan:2001kf,%
Feng:2001ib,Anchordoqui:2001ei,Uehara:2001yk}, 
for example, from interactions of
the cosmogenic neutrinos produced
in interaction of cosmic rays with the cosmic microwave background.
If interactions are not detected, then cosmic ray detectors could provide
constraints on the fundamental Planck scale for any number of extra
dimensions \cite{Ringwald:2001vk,Emparan:2001kf,%
Feng:2001ib,Anchordoqui:2001ei,Uehara:2001yk,Tyler:2000gt}. 

In this article we show that neutrino telescopes 
such as the Orbiting Wide-angle Light-collectors
Experiment (OWL) \cite{owl} 
and the Extreme Universe Space Observatory (EUSO)
\cite{Catalano:mm} have a very good chance of detecting black
holes produced in interactions of ultrahigh energy neutrinos from
extragalactic and cosmogenic sources and provide valuable information about the
fundamental Planck scale and the number of extra dimensions.  We
investigate whether the OWL neutrino telescopes can probe a region of parameter
space that is not accessible to LHC and Auger, and compare the reach
of OWL with a km$^3$ underground detector such as ICECUBE\cite{Halzen:1999jy}.

The OWL experiment will involve photodectectors mounted
on two satellites orbiting at 640 km above the Earth's surface. There are
three possible 
satellite configurations:a 500 km stereo view, a 
2000 km stereo view
and 2 monocular eyes. In case of the stereo configurations, 
the satellites are separated by a distance of 500 km or 2000 km respectively, 
such that they monitor a common region of the Earth's atmosphere 
\cite{jfk}. The 2000 km 
stereo configuration can view a larger volume but comparatively, the events 
are observed at a further distance from the satellites.
However, in the case of 2 monocular eyes configuration,
the satellites look down at the Earth's surface independently, therefore 
only one satellite will view an event. The EUSO experiment uses the same
principles as the OWL experiment, however, it is proposed to be 
a single eye located
on the international space station 380 km above the Earth's surface.
The geometric reduction in viewing volume, going from 2 eyes to one,
at a lower altitude, results in a reduction in the event rate by a factor of
$\sim 0.2$ compared to OWL. We will concentrate on OWL event rates below.

OWL detects ultrahigh energy neutrino interactions via air fluorescence. 
The large interaction lengths of neutrinos mean that
neutrinos initiate horizontal air showers. By
setting an angular threshold of column depth $>1500$ g/cm$^2$ (zenith
angle greater than $\sim 50^\circ$) in the atmosphere, neutrino
interactions are distinct from the hadronic and electromagnetic
showers initiated by cosmic rays \cite{owl}. Because
the electron in electron neutrino charged current interactions carries
a large fraction ($\sim 80\%$ of the incident neutrino energy) and
initiates an electromagnetic shower, detection of standard model
electron neutrino interactions is favored over muon neutrino
interactions. So that we can compare black hole event rates with the
standard model event rates, we concentrate on event rates initiated by
electron neutrinos and antineutrinos. We comment on the multiplicative
factors that are relevant when muon and tau neutrinos plus antineutrinos are
included.

Theoretical work has been done to set upper bounds on high energy
neutrino fluxes from active galactic nuclei (AGN) jets and 
gamma ray bursts (GRB) \cite{bounds}.  The bounds are
based on the theoretical correlations between the cosmic ray flux
and/or the extragalactic gamma ray flux and the neutrino flux.  These
bounds have some model dependence.  An upper bound, depending on cosmic
ray  source
evolution discussed recently by Waxman and Bahcall \cite{bounds} 
corresponds to the flux $dN_\nu/dE_\nu=1 \sim 5\times 10^{-8} 
(E_\nu/{\rm GeV})^{-2}$ 
(GeV\,cm\,s\,sr)$^{-1}$ for the
sum of muon neutrino plus antineutrino fluxes.
The electron neutrino plus antineutrino limits are a factor of $0.5$ lower
and are dominated by the neutrino component.
For convenience, we use 
\begin{equation}
dN_\nu/dE_\nu=10^{-8} (E_\nu/{\rm GeV})^{-2}
({\rm GeV\,cm\,s\,sr})^{-1}
\end{equation}
for the sum of the electron neutrino
plus antineutrino bound. In the case of bi-maximal $\nu_\mu\leftrightarrow
\nu_\tau$ neutrino oscillations, as favored by the SuperKamiokande
experimental data \cite{superk}, 
half of the muon neutrino plus antineutrino flux 
oscillates into tau neutrinos plus antineutrinos, yielding a 
ratio of 1:1:1 for electron, muon and tau neutrino flavors, each of which
we take as limited by Eq. (1).

\begin{figure}[!hbt]
\rule{0.0cm}{0.0cm}\\
\epsfxsize=7.5cm
\epsfbox[0 0  4096 4096]{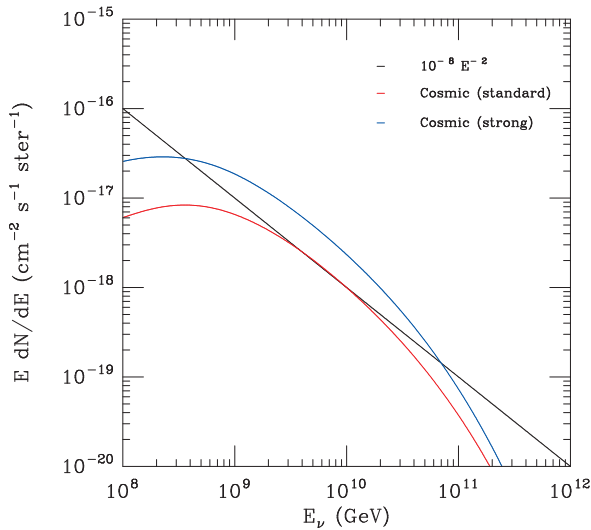}
\caption{The cosmogenic electron neutrino plus antineutrino flux evaluated
by Engel, Seckel and Stanev in [18].}
\end{figure}

While the sources of ultrahigh energy cosmic rays are not completely
understood, cosmic ray fluxes are measured. As the cosmic rays must traverse
a portion of the universe filled with the 3 K background radiation, at
sufficiently high energies, neutrinos will be produced by photoproduction
of charged pions which decay into neutrinos \cite{qj}. We use
here a new evaluation of this cosmogenic neutrino flux by
Engel, Seckel and Stanev \cite{ess}. They have evaluated the electron neutrino
plus antineutrino fluxes (and muon neutrino plus antineutrino fluxes) using
photoproduction rates based on the event generator SOPHIA \cite{sophia}.
They have presented results for two models of source evolution, one
with a parameterization scaling like $(1+z)^3$ for redshift $z<1.9$
(standard evolution), another scaling like $(1+z)^4$ for redshift
$z<1.9$ (strong evolution). Their results for the sum of the
electron neutrino and antineutrino fluxes,
for $E_\nu>10^8$ GeV, are shown in Fig. 1, 
along with a line associated with the approximate limit of Eq. (1).
The muon neutrino plus antineutrino flux is approximately a factor 
of two larger \cite{ess}.  The cosmogenic neutrino flux by Engel, Stecker 
and Stanev peaks at the same energy ($2-3\times 10^{8}$ GeV) 
as the flux calculated by Yoshida and Teshima \cite{yoshida} and 
by Protheroe and Johnson \cite{protheroe}.  The flux of Yoshida and 
Teshima is slightly narrower, while the agreement with the flux of 
Protheroe and 
Johnson is very good.

In the next section we review the black hole cross section. This is followed
by a discussion of the evaluation of the OWL event rates and a comparison
with the capabilities of ICECUBE. Our conclusions are presented in the final
section.

\section{Black Hole Cross Section}

At very high energies and at small impact parameters, in the case of
large extra dimensions,
neutrino-parton
interactions will result in a creation of a black hole.  For this to
happen, the center of mass energy has to be above the Planck scale
$M_D$, $\hat s\gg M_D^2$ and the impact parameter has to be much
smaller than the Scharzschild radius in $4+n$ dimensions.

In this case, the neutrino-parton cross section given by 
\begin{equation}
\hat\sigma (\nu j\rightarrow BH) = \pi r_S^2(M_{BH}=\sqrt{\hat s}) 
\theta (\sqrt {\hat s} - M_{BH}^{\rm min})$$
\end{equation}
where
$r_S$ is a Schwarzschild radius given by
\begin{equation}
r_S =\frac{1}{M_D}
\left[
\frac{M_{\rm BH}}{M_D}
\left(
\frac{2^n \pi^{\frac{n -3}{2}}\,\Gamma\left( \frac{3+n}{2}\right)}
{2+ n }
\right)
\right]^{\frac{1}{1+ n}}
\,.
\end{equation}
Here, $M_{\rm BH}^{\rm min}\gg M_D$ parameterizes the center
of mass energy above which the
semiclassical reasoning mentioned above is assumed to be valid.  It has been 
argued that in case of cosmic ray showers initiated by black hole decay one 
can relax this constraint because the details of the 
final state are not that important \cite{afgs}.  In our study, 
we will vary 
$M_{\rm BH}^{\rm min}$ from 1$\cdot M_D$ to 10$\cdot M_D$.  
We will consider $M_D\geq 1$ TeV for $n=2,4,6$. 

The neutrino-nucleon cross section for black hole production is given by 
\begin{equation}
\sigma (\nu N \to {\rm BH}) = 
\sum_i \int_{\frac{(M_{BH}^{\rm min})^2}{s}}^1 dx \,\, \hat\sigma_i^{BH}(xs)\,\, f_i(x,Q^2),
\end{equation}
where $s$ is the center of mass energy squared, $s=2m_NE_\nu$, 
and $f_i(x,Q^2)$
is the parton distribution function for parton $i$ \cite{cteq4}.
All partons contribute, and the antineutrino-nucleon cross section
for black hole production is identical to Eq. (4).

Qualitatively, we are interested in cross sections for $M_D\sim 1$ TeV, since
one motivation for TeV-scale compactification in the gauge hierarchy problem.
Scales much lower than 1 TeV would be manifest in collider experiments
by additional contributions from virtual graviton exchange, requiring
$M_D \gsim 1$ TeV \cite{afgs}. Direct graviton emission is already
constrained by LEP to $M_D>870$ GeV for $n=4$ and $M_D>610$ GeV for
$n=6$ \cite{1999jy,1999bz}.  
The radii of $n$ extra dimensions is of the order 
$R\sim 2\times 10^{-17}\,{\rm cm}\,
({\rm TeV}/M_D)\cdot(1.2\times 10^{16}\,{\rm TeV}/
M_D)^{2/n}$. 

\begin{figure}[!hbt]
\rule{0.0cm}{0.0cm}\\
\epsfxsize=7.5cm
\epsfbox[0 0  4096 4096]{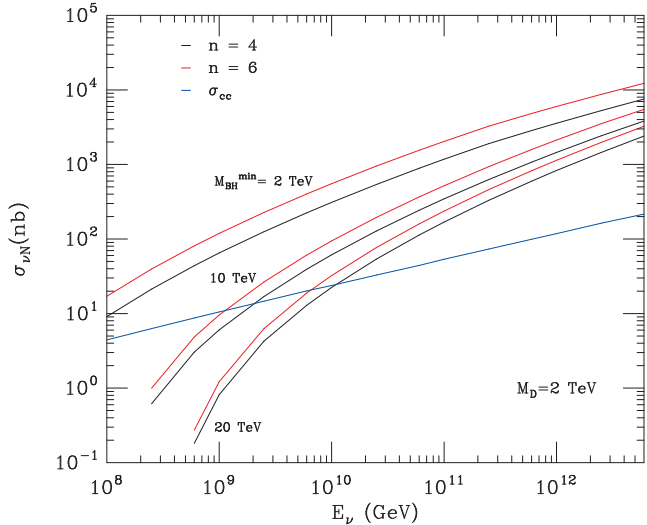}
\caption{Cross section for the black hole production in neutrino-nucleon 
scattering as a function of neutrino energy for $n=4,6$ and $M_{D}=2$ TeV.  
We also show the standard model charged-current cross section [28].  
}
\end{figure}

\begin{figure}[!hbt]
\rule{0.0cm}{0.0cm}\\
\epsfxsize=7.5cm
\epsfbox[0 0  4096 4096]{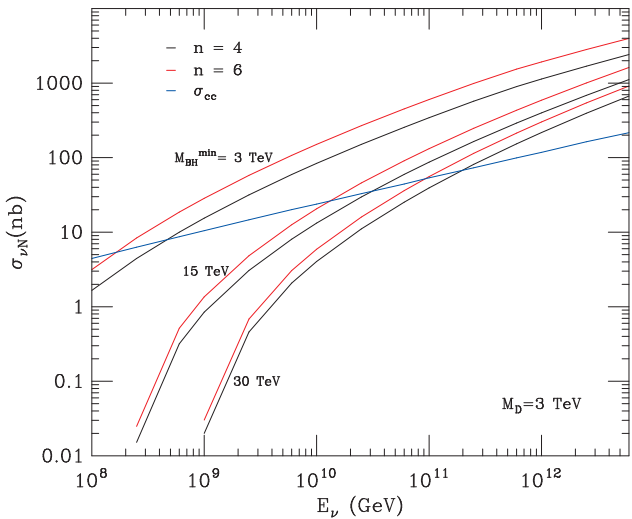}
\caption{Cross section for the black hole production in neutrino-nucleon 
scattering as a function of neutrino energy for $n=4,6$ and $M_{D}=3$ TeV. 
We also show the standard model charged-current cross section [28].  
}
\end{figure}

Deviations from Newtonian gravity over solar system distances exclude
the $n=1$ and  $M_D\sim 1$ TeV  possibility\cite{afgs}. For 
$n=2$, submillimeter tests of the gravitational inverse-square law constrain
$M_D > 3.5$ TeV \cite{2000cv} assuming two equal large extra dimensions.
For larger values of $n$, experimental non-observation of 
deviations from Newtonian gravity do not constrain $M_D>1$ TeV.
There are also astrophysical limits from, for example, supernova cooling
\cite{supernova}, however, these are more model dependent.

In Fig. 2, we show the cross sections for black hole production for
two values of the number of extra dimensions, $n=4,6$, for $M_D=2$
TeV as a function of incident neutrino energy for several values 
of $M_{BH}^{\rm min}$. We also show the standard model neutrino cross section
\cite{gqrs98}. For neutrino energies above $10^6$ GeV, $\sigma(\nu N)\simeq
\sigma(\bar{\nu}N)$ to within 5\%, so we need not distinguish between
incident
neutrinos and antineutrinos in our discussion. 
In Fig. 3, we show the same quantities, now with $M_D=3$ TeV. From these
plots, one sees that the standard model cross section is exceeded by the
black hole production cross section in  the energy range of $\sim 10^8-10^{11}$
GeV, depending on $n$, $M_D$, and $M_{BH}^{\rm min}$.
For $M_D=1$ TeV, the cross sections are even larger than in Figs. 2 and 3,
with the range of cross sections of $\sim 10-10^5$ nb for the same range
of energies when $M_{BH}^{\rm min}=M_D$.

>From Figs. 2 and 3, we see that for the energies of interest for the
OWL detector, the largest black hole
production cross sections are  on the order of $10^4$ nb. While these are 
larger than the standard model neutrino-nucleon cross sections, they are 
still small compared to typical strong interaction cross sections
in the range of tens of millibarns. As a result, even with the enhanced
cross sections for black hole production, the neutrino still penetrates
deep into the atmosphere, as demonstrated below.

\section{Detection of Black Hole Production with OWL}

The detection of neutrino production of black holes by interactions of
neutrinos with nuclei in the atmosphere follows the same principle
as detection of neutrinos via their standard model interactions.
Since neutrinos are weakly interacting, they are more likely to 
penetrate the atmosphere in the horizontal direction, whereas cosmic
rays interact in a shell about 20 km above the surface of the Earth.

To estimate the critical cross section, below which the probability of
interaction in the atmosphere is peaked at sea level, one can 
compare the column depth of the atmosphere with the interaction
cross section. The column depth at zenith angle $\theta$ is
\begin{equation}
X=\int_0^\infty dx \rho(h(x,\theta))
\end{equation}
as measured along the particle trajectory from a point on the surface
of the Earth (of radius $R_\oplus$),
in terms of the atmospheric density $\rho$ as a function of
altitude $h=\sqrt{R_\oplus^2+2xR_\oplus\cos\theta +x^2}-R_\oplus$. 
To a good approximation, the US Standard Atmosphere (1976)
\cite{atm} is
\begin{equation}
\rho_{atm}(h)=\cases{1.225\times 10^{-3} \ {\rm g/cm}^3\exp(-h/9.192\ {\rm km})
, & \cr
\quad\quad h<10\ {\rm km} & \cr
1.944\times 10^{-3} \ {\rm g/cm}^3\exp(-h/6.452\ {\rm km})
, &\cr
\quad \quad  h\ge 10\ {\rm km}\ .& \cr} 
\end{equation}
Numerically, the column depth for neutrinos arriving vertically is $10^3$ 
cmwe, while the column depth for neutrinos arriving horizontally
is $3.6\times 10^4$ cmwe.
By comparison, the neutrino interaction length is
\begin{equation}
\lambda_\nu = 1.7\times 10^9\cdot (\sigma/{\rm nb})^{-1}\ {\rm cmwe}\ ,
\end{equation}
so for $\sigma<\sigma_{crit}\simeq 5\times 10^4$ nb, the horizontal
column depth is larger than the interaction length.
The black hole cross section in Figs. 2 and 3
is below $10^4$ nb for $E_\nu<10^{12}$ GeV, so
$X/\lambda_\nu<0.2$ for horizontal neutrinos. In fact, 
the cosmogenic neutrino flux falls with energy more rapidly than
$E_\nu^{-2}$ above $E_\nu\sim 10^8$ GeV, and the Waxman-Bahcall bound falls
with $E_\nu^{-2}$, so
the bulk of the 
contribution to the event rate is at  $E_\nu\ll 10^{12}$ GeV where $X/\lambda_\nu
\ll 1$. 

\begin{figure}[!hbt]
\rule{0.0cm}{0.0cm}\\
\epsfxsize=7.5cm
\epsfbox[0 0  4096 4096]{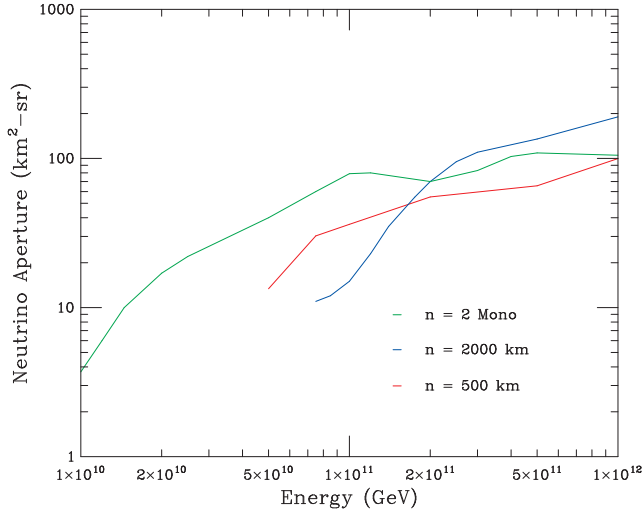}
\caption{OWL electron neutrino (standard model) effective
aperture as a function of energy for
$500$ km, $2000$ km satellite separation and 2 monocular eyes [11].
}
\end{figure}

The standard model electron neutrino effective aperture is shown in Fig. 4
\cite{owl}.
At high energies, the effective aperture roughly scales with energy as the
standard model neutrino-nucleon cross section does. 
To a good approximation, all of the energy of the
decaying black hole is deposited into hadronic 
or electromagnetic showers, just as all of the energy in electron neutrino
charged current interactions is deposited in the shower.
Since the neutrino interaction length, even when
black hole production is included, is small, one may evaluate the 
number of black holes detected with OWL by rescaling the neutrino aperture
for electron neutrinos by the ratio of the black hole cross section to
the neutrino-nucleon standard model charged current cross section
($\sigma_{CC}^{SM}$):
$$
N  = T { \int_{\small{E_{min}}}^\infty
\epsilon A_{\small Nuc}(E){\sigma_{BH}(E_\nu)\over
\sigma_{CC}^{SM}(E_\nu)}\frac{dN_{\nu}}{dE_{{\nu}}}}dE_\nu\ ,
$$
where $T $ is the duration of data taking, $\epsilon = 0.1 $ is the
duty cycle, $A_{\small Nuc}(E)$ is the OWL aperture as it appears
in Fig. 4 \cite{owl},
${dN_{\nu}}/{dE_{\nu}}$ is the neutrino flux and
$\sigma_{BH}(E_\nu)$ is the cross section for the production of black
hole.

\begin{figure}[!hbt]
\rule{0.0cm}{0.0cm}\\
\epsfxsize=7.5cm
\epsfbox[0 0  4096 4096]{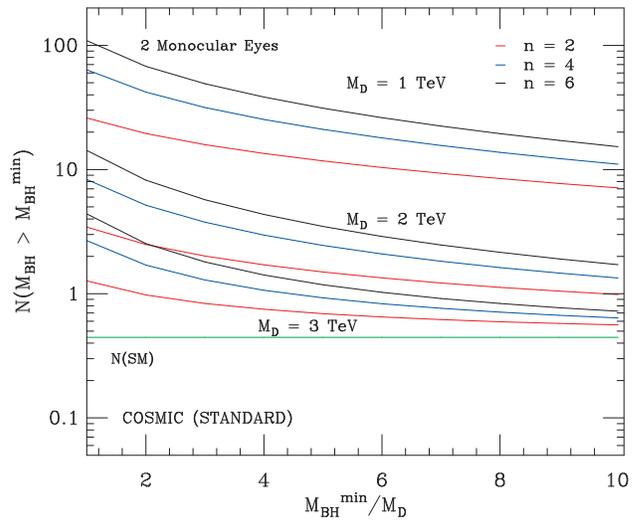}
\caption{Event rate per year for black hole production 
plus standard model background as a function of 
$M_{BH}^{\rm min}/M_{D}$ for cosmic (standard) incident $\nu_e+\bar{\nu}_e$
flux. The OWL rates are shown in case of 2 monocular eyes for 
$n=2,4,6$ extra dimensions and $M_{D}=1,2,3$ TeV. 
We also show separately the standard model event rates.}
\end{figure}

The resulting event rates are shown in Figs. 5 and 6 for the standard evolution
and the strong evolution $\nu_e+\bar{\nu}_e$ cosmogenic fluxes of
Engel, Seckel and Stanev \cite{ess}.
We show the results for the 2 monocular eyes configuration. The stereo 
configurations
have larger threshold energies which gives smaller event rates.  
The standard model contributions are also shown.
In the case of the cosmogenic electron neutrino plus antineutrino
flux obtained with the 
standard evolution, the background from the standard model 
charged current interaction is very small (about 0.4 events per year), 
while the showers from black hole evaporation give two 
events 
per year for $M_D=M_{BH}^{\rm min}=3$ TeV and 
$n=4$, and gives between three and eight 
events for $M_D=2$ TeV, $M_{BH}^{\rm min}\leq 4 M_D$ and $n=4$.  
As many as 100 events
are possible with $M_D=1$ TeV and $n=6$.
The cosmogenic neutrino 
flux  with strong evolution gives larger event rates, by about a factor of 
2, both for the signal and the background.  

\begin{figure}[!hbt]
\rule{0.0cm}{0.0cm}\\
\epsfxsize=7.5cm
\epsfbox[0 0  4096 4096]{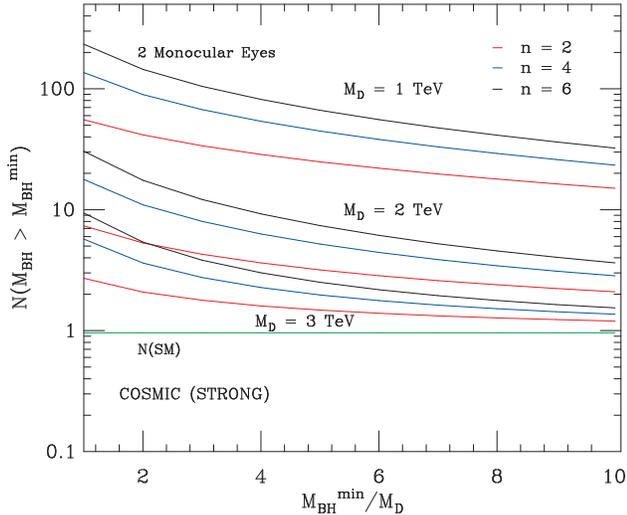}
\caption{Event rate per year for black hole production plus standard
model background as a function of 
$M_{BH}^{\rm min}/M_{D}$ for cosmic (strong) incident $\nu_e+\bar{\nu}_e$
flux. The OWL rates are shown in case of 2 monocular eyes for 
$n=2,4,6$ extra dimensions and $M_{D}=1,2,3$ TeV. 
We also show separately the standard model event rates.}
\end{figure}

\begin{figure}[!hbt]
\rule{0.0cm}{0.0cm}\\
\epsfxsize=7.5cm
\epsfbox[0 0  4096 4096]{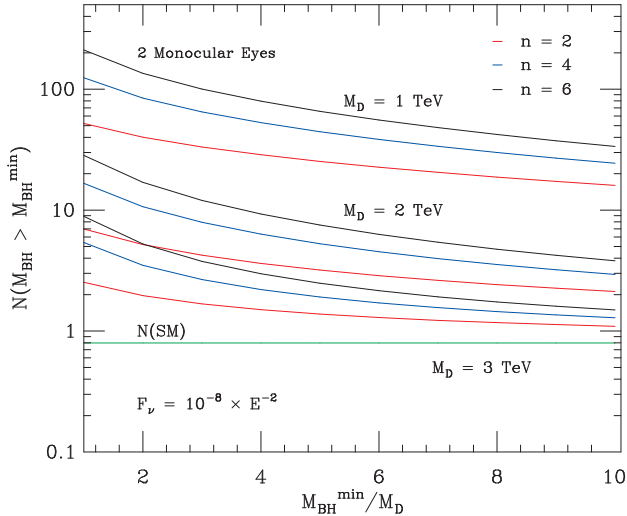}
\caption{Event rate per year for black hole production plus standard
model background as a function of 
$M_{BH}^{\rm min}/M_{D}$ 
for $dN_\nu/dE_\nu=10^{-8}E_\nu^{-2}$ incident neutrino
flux. The OWL rates are shown in case of 2 monocular eyes for 
$n=2,4,6$ extra dimensions and $M_{D}=1,2,3$ TeV. 
We also show separately the standard model event rates.}
\end{figure}

Similar results for event rates are obtained with the Waxman-Bahcall bound
on the neutrino flux from optically thin sources as represented
by Eq. (1). These are shown
for the 2 monocular eyes in Fig. 7.  For $M_D=1$ TeV and $n\geq 4$, 
we find that OWL would detect between 20 and 200 events per year, 
about two orders of magnitude above the standard model predictions.  
For larger $M_D$, the  rates decrease so that for $M_D=M_{BH}^{\rm min}=2$ 
TeV and 
$n\geq 4$, there are tens of events per year.  
For $M_D=M_{BH}^{\rm min}=3$ 
TeV and $n \geq 4$ there are handful of events with 
background of 0.8 events.

The event rates above can be compared with contained rates for electron
neutrino interaction in a km$^3$ detector like ICECUBE. 
The contained event rate for black hole production is
$$
{N\over T} =  \int dE_{\nu_\mu} (V_{eff})^{2/3}
\:\Biggl(1-\exp \bigl[-z N_A\sigma_{BH}(E_{\nu})\bigr]\Biggr)
 \frac{dN_{\nu}}{dE_{{\nu}}} .
$$
where $N_A = 6.022 \times 10^{23}$ is Avogadro's number and $
{dN_{\nu}}/{dE_{\nu}} $ is the neutrino flux that reaches the
detector of volume $V_{eff}$, and accounting for the large scale of the
detector, $z=(V_{eff})^{1/3}\rho_{ice}$.  We take $V_{eff} =1$ km$^3$
corresponding to the size of the planned neutrino detector ICECUBE.
The contained event rates appear in Tables I-III 
using an energy threshold
of $10^{8}$ GeV for $M_D=1,2,3$ TeV. 
For $M_D= 1$ TeV, there are a handful of event in ICECUBE, depending on the 
value of the number of extra dimensions, for all three fluxes.
The standard model rates per year are 0.08 (0.03, 0.09) for the 
Waxman-Bahcall bounded $E^{-2}$ (standard cosmogenic, strong cosmogenic)
flux with the same energy threshold.
Our rates agree qualitatively with the recent results of Alvarez-Mu\~niz 
{\it et al.} \cite{fenghalzen}, who used
a lower threshold energy, included $\nu_\mu$ and $\nu_\tau$ fluxes,
and had different assumptions about the shower energy in the decay
of the black hole. We comment on the consequences of these different
assumptions for the OWL rates in the next sections.
  
The OWL rates as shown in the Figures 
are a factor of more than 20 times higher than the ICECUBE rates
in Tables I-III.  Lowering the energy threshold for ICECUBE 
increases the event 
rates for the Waxman-Bahcall flux bound,
but decreases the signal to 
background ratio.  The cosmogenic neutrino flux is less sensitive
to the energy threshold because it does not fall $\sim E_\nu^{-2}$ for
$E_{\nu}<10^8$ GeV.
ICECUBE should be able to 
detect black hole events (showers and muons) with $E_{th} \sim 10^5$ GeV 
if $M_D \leq 2$ TeV and $n \geq 6$ \cite{fenghalzen}, however, 
at higher energies OWL will have
more sensitivity.

\section{Discussion}

The event rates in Figs. 5-7 and in the Tables include only incident
electron neutrino and antineutrino fluxes.
Since the cross section for neutrino production of black holes is lepton
flavor-blind, the black hole production rates can be multiplied by the
ratio of the total neutrino plus antineutrino flux to the electron
neutrino plus antineutrino flux. The net effect is essentially a factor
of three, since generically, the neutrinos are coming from $\pi^+\rightarrow
\nu_\mu \mu^+\rightarrow \nu_\mu \bar{\nu}_\mu \nu_e e^+$. This is borne
out numerically in, for example, the cosmogenic flux of Engel, Seckel
and Stanev. This leads to a factor of three enhancement of all of the
black hole rates in Figs. 5-7 and in Tables I-III. 
The standard model rates are not necessarily increased by
such a large factor because the charged lepton carries,
on average, about 80\% of the incident neutrino energy after the
charged current interaction, leaving the shower with 20\% of the incident
neutrino energy. For a flux falling like $E_\nu^{-2}$
and standard model cross sections that increase
like $\sim E_\nu^{0.4}$ \cite{gqrs98}, the rate of hadronic
showers from the charged current interaction will be suppressed by
a factor of $\sim (0.2)^{0.6}=0.4$. By including all three neutrino flavors,
the standard model rates increase by roughly a factor of 2. 
For the contained rates induced by cosmogenic
neutrinos with a minimum shower energy around $10^8$ GeV,
the standard model rates are increased by a factor of $\sim 3$
for the total neutrino flux. This is due to the relative
insensitivity of the event rate
to the minimum shower energy with the flatter shape of 
the cosmogenic $E_\nu dN_\nu/dE_\nu$ below $E_\nu\sim 10^9$ GeV.

A related issue is the uncertainty in the event rates presented here
due to the approximation that all of the black hole energy goes into the
showers. It has been argued that only $\sim 75\%$ of the incident neutrino
energy ends up in the hadronic shower associated with the black hole decay
\cite{fenghalzen}.
Because the black hole cross sections grow with incident neutrino energy
like $\sim E_\nu^{0.5}-E_\nu^{0.8}$, 
given $dN_\nu/dE_\nu\sim E_\nu^{-2}$ and $E_{shr}=0.75 E_\nu$,
the rates will be suppressed by $(0.75)^{0.2}-(0.75)^{0.5}$, between
a 5-15\% decrease in the rates. The suppression will be even less for
the contained  cosmogenic rates since the shower threshold dependence
is weak for $E_{shr}^{min}\sim 10^8$ GeV.

Neutrino fluxes that consistently decrease like $E_\nu^{-2}$ will most
likely be seen at lower energy thresholds than
considered here, in detectors like ICECUBE.
Alvarez-Mu\~niz {\it et al.} in Ref. \cite{fenghalzen} have emphasized  
the different (and complementary) signals of muons, taus and showers
that can be used as diagnostics. For the cosmogenic flux, the spectrum
is such that higher energies are emphasized and larger volumes are required,
as seen by a comparison of Figs. 5-7 and the tables. 

The OWL telescope has the capability of probing the
cosmogenic flux farther than ground-based
air shower experiments.
The non-observation of an excess of
shower events at the AGASA air shower array lead to limits on the
black hole production parameters and require $M_D \geq 1.3 -
1.8$ TeV \cite{afgs}. 
The OWL standard model rates are at the 1.5-3 events per year level,
depending on the flux, when one 
includes all flavors of neutrinos. 
The rates for OWL are ten to hundreds of events per year for 
$M_D=1$ TeV for $M_{BH}^{\rm min}=1-10$, even in the case of the 
conservative evolution of 
the cosmogenic flux. Similar results are found for the Waxman-Bahcall flux,
which represents the upper bound for optically thin sources.
Multiplying the black hole rates for cosmogenic fluxes in the
figures by a factor of three means that for $M_{BH}^{min}=5\, (1)\, M_D$, 
$M_D=3$ TeV, $n=6$,
the  annual signal event rates will be on the order of 4.4 (24)
for the strong
evolution cosmogenic flux model and 2.4 (12) for the standard evolution.
This is a much larger reach in parameter space that the terrestrial
experiments. One year of
data taking would be sufficient for OWL to have unique opportunity to
detect black holes, or to probe fundamental Planck scale up to
$M_D=3$ TeV for $n\geq 4$. OWL is set for possible implementation
after 2007. EUSO is proposed to go on the International Space Station
in 2006.
EUSO, with a projected event rate on the order of 1/5 of the OWL rate,
will be able to probe regions of parameter space intermediate between
ICECUBE and OWL.

\vskip 0.1true in

\leftline{\bf Acknowledgments}

\vskip 0.1true in

The work of S.I.D. has been supported in part by National Science Foundation Grant
0070998. The work of I.S. has been supported in part by the DOE under
Contracts DE-FG02-95ER40906 and DE-FG03-93ER40792.  The work of
M.H.R. has been supported in part by National Science Foundation Grant
No.  PHY-9802403. We thank J. Krizmanic for discussions and providing
the neutrino acceptance for OWL. We thank T. Stanev for providing data
files with the ESS fluxes.

\begin{table}
\caption{Downward 
contained event rates per year for $M_{D}$ of 1 TeV $(M_{BH}^{\rm min}=
m$ TeV).}
\begin{centering}
\vspace{0.2cm}
\end{centering}
\begin{tabular}{lcccc}
$n$ & $m$ &  $E^{-2}$  & $ COS_{STD} $ & $ COS_{STR} $ \\
2 & 1 & 1.58    & 0.79 & 2.13 \\
2 & 5 &  0.36   & 0.22 & 0.56 \\
2 & 10 &  0.15  & 0.10 & 0.24 \\
4 & 1  & 4.44  & 2.16 & 5.89 \\
4 &  5 &  0.72  & 0.43 & 1.11 \\
4 & 10 &  0.24 & 0.16 & 0.40 \\
6 & 1  & 7.98  & 3.85 &10.50 \\
6 & 5  & 1.11 & 0.66 & 1.71 \\
6 &10  & 0.35 & 0.24 & 0.58 \\
\end{tabular}
\end{table}

\begin{table}
\caption{Downward 
contained event rates per year for $M_{D}$ of 2 TeV $(M_{BH}^{\rm min}=m$
TeV).}
\begin{centering}
\vspace{0.2cm}
\end{centering}
\begin{tabular}{lcccc}
$n$ & $m$ & $E^{-2}$  & $ COS_{STD} $ & $ COS_{STR} $ \\
2&2 &0.15  &0.078 &0.207\\
2&10  & 0.023  &0.015 &0.038\\
2&20  & 0.0084  &0.005 &0.013\\
4& 2  &0.43  &0.225 &0.599\\
4&10  & 0.046  &0.031 &0.077\\
4&20  & 0.014  &0.009 &0.022\\
6& 2  &0.78 &0.406 &1.085\\
6&10  & 0.072  &0.048 &0.119\\
6&20  & 0.020  &0.014 &0.032\\
\end{tabular}
\end{table}

\begin{table}
\caption{Downward 
contained event rates per year for $M_{D}$ of 3 TeV $(m=M_{BH}^{\rm
min}/M_{D})$.}
\begin{centering}
\vspace{0.2cm}
\end{centering}
\begin{tabular}{lcccc}
$n$&$m$  & $E^{-2}$  & $ COS_{STD} $ & $ COS_{STR} $ \\

2&3 & 0.034   & 0.019 & 0.050\\
2&15 & 0.0043 & 0.0029 & 0.0069\\
2&30 & 0.0016 & 0.00091 & 0.0020\\
4&3  &0.10      &0.056 &0.15\\
4&15  & 0.0088  &0.0060 &0.014\\
4&30 & 0.0026   &0.0016 &0.0036\\
6&3  &0.19 &0.10 &0.27\\
6&15  & 0.014 &0.0094 &0.022\\
6&30  & 0.0038 &0.0023 &0.0053\\
\end{tabular}
\end{table}

\end{document}